\documentclass[aps,prl,twocolumn,showpacs]{revtex4}
\usepackage{bm}
\usepackage{graphicx}
\usepackage{amsmath}
\usepackage{eufrak}
\usepackage{color}
\newcommand{\nix}[1]{}
\begin{document}

\title{Photogalvanic effects due to quantum interference
in optical transitions demonstrated by terahertz radiation absorption in
Si-MOSFETs}

\author{P. Olbrich,$^1$  S.A. Tarasenko,$^2$ C.~Reitmaier,$^1$
J.~Karch,$^1$ D. Plohmann,$^1$  Z.D. Kvon,$^3$ and
S.D.~Ganichev$^{1}$}
%
\affiliation{$^1$  Terahertz Center, University of Regensburg, 93040
Regensburg, Germany,}
\affiliation{$^2$A.F.~Ioffe Physico-Technical Institute, Russian
Academy of Sciences, 194021 St.~Petersburg, Russia}
\affiliation{$^3$ Institute of Semiconductor Physics, Russian Academy
of Sciences, 630090 Novosibirsk, Russia}

\begin{abstract}
We report on the observation of the circular (helicity-dependent)
and linear photogalvanic effects in Si-MOSFETs with
inversion channels.
%
%
%
The developed microscopic theory demonstrates that
the circular photogalvanic effect in Si structures it is of pure
orbital nature originating from the quantum interference of
different pathways contributing to the light absorption.
\end{abstract}

\date{\today}

\maketitle


The  photogalvanic
effects have been proved to be an efficient tool to study
nonequilibrium processes in semiconductor structures yielding
information on their
%
%
symmetry, details of
the band spin splitting, momentum, energy and spin relaxation
times  etc. (see e.g.\,\cite{SturmanFridkin_book,Reimann02,Ivchenko_book,GanichevPrettl_book}).
%
%
Microscopically,  they
%
%
are caused by the asymmetry of photoexcitation or
relaxation processes and, thus, can be observed in a media of
sufficiently low spacial symmetries only.
%
%
The photogalvanic effects
%
%
have recently been observed  in a large class of low-dimensional
structures for interband\,\cite{Belkov03,Bieler05,Yang06,Cho07}
and intraband  optical
transitions\,\cite{GanichevPrettl_book,Ganichev01,Ganichev03}.
%
%

Here we report on the observation of the circular (CPGE)
and linear (LPGE) photogalvanic effect
caused by
absorption of THz radiation in Si-MOSFETs.
%
%
%
The fact of the existence of the
helicity-sensitive circular photogalvanic current, which reverses
its direction upon switching the sign of circular polarization,
in Si-based structures is of particular interest. So far, the CPGE
has only been detected in materials with strong spin-orbit
coupling and described by microscopic mechanisms based on
spin-related processes\,\cite{Ivchenko_book,GanichevPrettl_book}.
However, such mechanisms of the CPGE get ineffective in Si
%
%
because of the vanishingly small constant of spin-orbit coupling
and, therefore, can not account for the observed circular
photocurrents. Thus, a new access in explaining the CPGE is
required involving mechanisms of pure orbital (spin-unrelated)
origin. 
%
Here, we show that the CPGE in
our structures is due to
%
%
quantum interference of different pathways contributing to light
absorption\,\cite{Tarasenko07} (see
also\,\cite{Magarill89,IvchenkoPikus_book}). In contrast to the
well known photocurrents caused by all-optical quantum
interference of one- and two-photon absorption processes in a two
color light\,\cite{Entin89,Hache97,Bhat00p5432,Stevens02p4382},
here the elementary one-photon absorption processes give rise to
an electric current. Beside the CPGE, we also
investigate the LPGE, which has so far been detected in Si-MOSFETs
for direct intersubband transitions only~\cite{Gusev87}. Our
results show that the  free-carrier
absorption also leads to the LPGE.
%
%
%

\begin{figure}[t]
\includegraphics[width=0.75\linewidth]{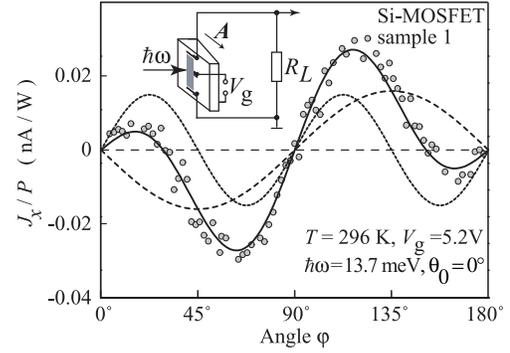}
\caption{Normalized photocurrent $J_x/P$ measured in sample\,1 a function of the angle $\varphi$.
Full line is a  fit to  Eqs.\,\protect(\ref{phi-comp}). Dashed and doted lines show
the CPGE and LPGE contributions, respectively.
The inset sketches the experimental set-up.}
\label{figure1phiRT90mkm}
\end{figure}

We study $n$-type MOSFETs
prepared on miscut
%
%
Si surfaces.
%
%
The surfaces  of our samples are
tilted by the angle $\theta=9.7^\circ$ (sample\,1) or $\theta=10.7^\circ$
(sample\,2) from the $(001)$
plane around  $x \parallel [1\bar{1}0]$. Two transistors
oriented along and normal to the inclination direction $\bm
A \parallel y$ with  semitransparent Ti gates of $10$\,nm thickness
are prepared on each substrate. 
%
%
%
 Application of the gate voltage $V_g = 1$ to
$10$\,V enables us to change the carrier density $N_s$ and the
energy spacing $\varepsilon_{21}$ between the size-quantized subbands $e1$ and $e2$
in the range of $N_s = 1.5$ to $15 \times
10^{11}$\,cm$^{-2}$ and $\varepsilon_{21}=2$ to $20$\,meV,
respectively.  The peak electron mobility $\mu$ in the
channel
is about 10$^3$ and $2\times 10^4$\,cm$^2$/Vs at
$T = 296$ and
$4.2$\,K.

For optical excitation we applied THz radiation of a
pulsed  optically pumped NH$_{3}$ laser\,\cite{GanichevPrettl_book}.
The laser generates radiation pulses with a power $P \simeq $ 5\,kW and
wavelengths $\lambda =$\,76, 90.5, and 148\,$\mu$m corresponding to
the photon energies $\hbar\omega =$ 16.3, 13.7 and 8.4\,meV,
respectively.
%
%
%
Quartz  $\lambda/4$ and $\lambda/2$
plates were used to modify  the laser light polarization.
Rotating the $\lambda/4$ plate by the angle $\varphi$ between
the plate optical axis and the incoming laser polarization
we varied the radiation helicity as $P_{\rm circ} = \sin{2 \varphi}$.
%
%
By means of the $\lambda/2$ plates we obtain the
linearly polarized light with all possible angles $\alpha$ between
the electric field of radiation and the $x$ axis.
%
The photocurrents are measured between
source and drain
via the voltage drop across a 50\,$\Omega$ load resistor.
%
%

\begin{figure}[t]
\includegraphics[width=0.65\linewidth]{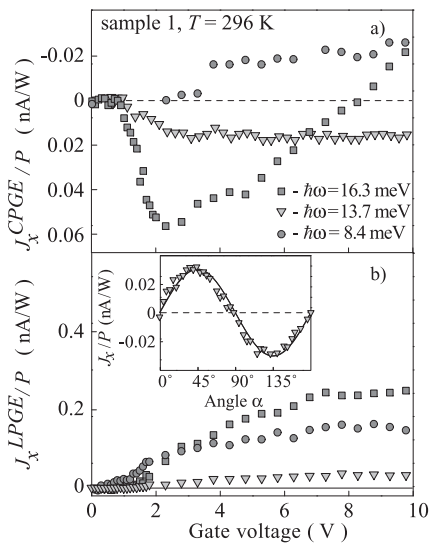}
\caption{Normalized photogalvanic currents as a function of the gate voltage.
a) CPGE,
b)
LPGE
The inset  shows the LPGE current
as a function of the azimuth angle $\alpha$.} \label{figure2_gateRTbis}
\end{figure}

Irradiating MOSFET structures by polarized light at normal
incidence, as sketched in the inset to
Fig.\,\ref{figure1phiRT90mkm}, we observed a photocurrent signal
with the temporal structure reproducing the laser pulse of about
100\,ns duration.
The signal is detected only for the gate voltages in
the range of $V_g = 1$ to $10$\,V.
%
%
%
%
%
First we discuss data obtained at room temperature. Applying
radiation to the transistors aligned
perpendicular to $\bm A$ (see the inset in Fig.\,\ref{figure1phiRT90mkm}) and
varying the radiation helicity, we observe that the induced photocurrent
%
%
reverses its direction upon switching the radiation helicity
from left- to right-handed circular polarization. In
contrast, the photocurrent  measured in transistors
aligned along $\bm A$
is observed to be the same for both left- and right-handed circularly
polarized light. The fact that the CPGE in miscut structures at normal incidence is
observed only for the direction normal to $\bm A$   is in accordance
with the phenomenological theory of the photogalvanic effects
structures of the $C_s$ point-group symmetry
relevant to our miscut samples\,\cite{Ganichev01}.
%
%
Our work is mostly devoted to the CPGE, therefore  we focus below on this type of transistors only.
Figure\,\ref{figure1phiRT90mkm} shows the dependence
of the photocurrent on the angle $\varphi$ for radiation with $\hbar\omega = 13.7$\,meV
measured for sample\,1.
The data
can be well fitted by the equation\,\cite{Ivchenko_book,GanichevPrettl_book}
\begin{eqnarray} \label{phi-comp}
J(\varphi) = J_{C} \sin{2 \varphi} + (J_{L}/2) \sin{4\varphi} \:.
\end{eqnarray}
%
%
Here, the first and the second terms
%
%
describe the CPGE and LPGE contribution to the photocurrent.
%
%
Figure\,\ref{figure1phiRT90mkm} shows substantial contribution of
the CPGE to the total current.
%
This feature persists for all gate voltages
%
%
and all photon energies used in our experiments.

\begin{figure}[t]
\includegraphics[width=0.6\linewidth]{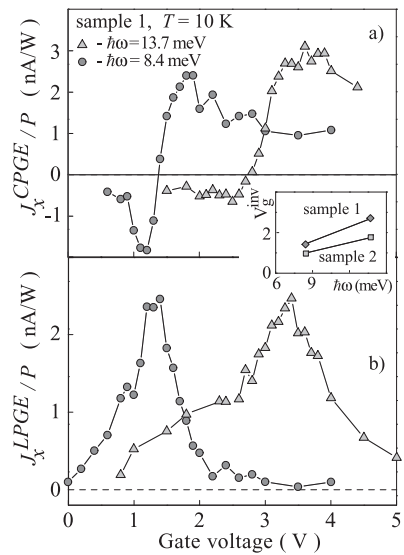}
\caption{Normalized circular photogalvanic current (a)
and linear photogalvanic current (b)
as a function of the gate voltage.
The inset shows the  spectral dependence of the gate voltage corresponding to the
sign inversion of the CPGE.
} \label{figure3_gateLowT}
\end{figure}

Figure\,\ref{figure2_gateRTbis}a shows the dependences of the CPGE
contributions $J_{C}$ to the total photocurrent measured   as a
function of the gate voltage $V_g$.
The CPGE contribution is obtained by taking the difference between
photoresponces to the right- and left-handed radiation yielding the CPGE current
$J_{C} = [J(\varphi = 45^\circ) - J(\varphi = 135^\circ)]/2$.
%
%

To investigate the LPGE contribution in more detail, we excited our
samples with linearly polarized radiation which excludes the CPGE.
%
%
The photocurrent can be well fitted by the
phenomenological equation\,\cite{GanichevPrettl_book}
\begin{eqnarray} \label{alpha-comp}
J(\alpha) =  J_{L} \sin{2\alpha} \:,
\end{eqnarray}
with the same parameter $J_{L}$ as used in  Eq.\,(\ref{phi-comp}).
The dependence of the LPGE contribution on the gate voltage is shown
in Figure\,\ref{figure2_gateRTbis}b where $J_{L} = [J(\alpha = 45^\circ) - J(\alpha = 135^\circ)]/2$
is plotted for several radiation photon energies.
%

As follows from Fig.\,\ref{figure2_gateRTbis}b,
the LPGE current has
the same sign for all radiation
photon energies used and its magnitude increases with the gate
voltage.  Such a dependence can be attributed to the increase of the electron density
in the inversion channel and, therefore, the Drude absorption
enhancement. The CPGE behaviour, in contrast, is more complicated (see Fig.\,\ref{figure2_gateRTbis}a).
We observe that the direction of the CPGE current is opposite for
$\hbar \omega$= 8.4\,meV and 13.7\,meV. Moreover, for $\hbar \omega$= 16.3\,meV
the signal changes its sign with the  bias voltage increase. We note that we can not
attribute this gate voltage to any characteristic energy in the band structure.

At low temperatures the behaviour of the photocurrent
upon variation of the radiation polarization remains the same.
All data can be well fitted by Eqs.\,(\ref{phi-comp}) and\,(\ref{alpha-comp}) with comparable magnitudes
of $J_C$ and $J_L$ proving the presence of both, CPGE and LPGE.
The gate voltage behaviour, however,
changes drastically as demonstrated in Fig.\,\ref{figure3_gateLowT}a
for sample\,1. The LPGE current, instead of smooth dependence observed at room
temperature, shows a resonant response (see Fig.\,\ref{figure3_gateLowT}b).
The peak position depends on the photon energy and corresponds to
the $\hbar\omega \cong \varepsilon_{21}$
subband separation. The resonance is obtained by tuning the band separation
to the photon energy varying the gate voltage.
Increasing of the photon energy should shift
the intersubband resonance  to the large gate voltages\,\cite{Ando82}.
The position of the inter-subband resonance has been additionally proved
by the photoconductive measurements in biased transistors.
%
%
Note, that the difference in the
resonance position for fixed photon energy observed in
sample 1 and 2 is attributed to the difference in the declination angle.
In the CPGE we observe again the change of the current direction but, now,
at all frequencies used. In contrast to the  room temperature data,
the gate voltage inversion at low temperatures takes place at the
resonance $\hbar\omega \cong \varepsilon_{21}$. In fact,
the CPGE photocurrent changes its sign upon the gate voltage variation
and vanishes at the gate voltage values at which the LPGE achieves its maximum
(see Fig.\,\ref{figure3_gateLowT}). The inset of Fig.\,\ref{figure3_gateLowT}
shows the gate voltages of the CPGE inversion  as a function of the photon energy.
\begin{figure}[t]
\includegraphics[width=0.97\linewidth]{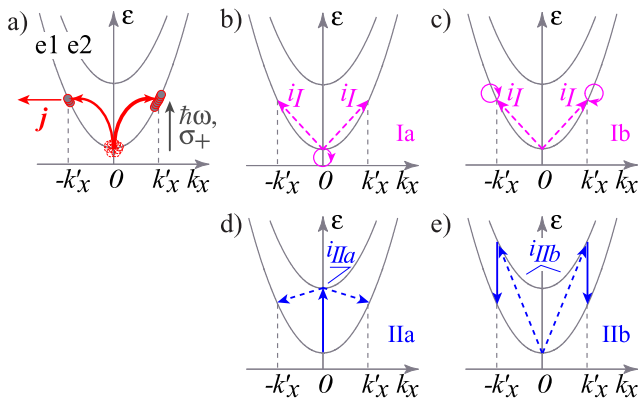}
\caption{Microscopic model of the CPGE. a) Indirect 
optical transitions  due to absorption of circularly
polarized light are shown
by bend arrows of various thickness indicating the difference in
transition rates for the absorption of circularly polarized
radiation caused by the quantum interference of various pathways.
Circles sketch the resulting imbalance of the carrier distribution
yielding an electric current $\bm{j}$. b) -- e) 
Various pathways via intermediate states in the $e1$ and $e2$
subbands. Here, solid arrows indicate electron-photon interaction
and the dashed arrows describe
%
%
scattering events. 
} \label{figure4_model}
\end{figure}

The observation of the CPGE
%
%
%
%
apart the inter-subband resonance demonstrates that the free
carrier absorption of circularly polarized light gives rise to the
helicity dependent current.
Below we consider theoretically this
process
%
%
and show that the CPGE is caused by the interference of different pathways
contributing to the radiation absorption. Figure\,\ref{figure4_model}a
sketches the indirect optical transitions
within the ground subband $e1$.
%
%
%
Due to the energy and momentum conservation, the transitions shown
by bend arrows can only occur if the electron-photon interaction
is accompanied by simultaneous electron scattering by phonons or
static defects. Such optical transitions involving both the
electron-photon interaction and electron scattering are treated as
second-order virtual processes via intermediate states.
%
Figure\,\ref{figure4_model}b -- d shows possible absorption pathways
%
%
with the intermediate states
%
%
in the
$e1$ and
$e2$
subbands.

The dominant contribution to the Drude absorption
involves intermediate states within the  $e1$ subband.
 Such transitions (path I) are shown in
Figs. \ref{figure4_model}b,\,d for the process where
the electron-photon interaction is followed by electron
scattering (Fig.\,\ref{figure4_model}b) and the inverted sequence
process (Fig.\,\ref{figure4_model}d). The matrix element of the
intrasubband optical transitions with intermediate states in the
$e1$ subband on the vicinal silicon surface has the
form\,\cite{IvchenkoPikus_book,Ando82}
\begin{equation}\label{Me1}
M_{\mathbf{k}'\mathbf{k}}^{(1)} = \frac{eA}{c \omega} \left[
\frac{(k_x'-k_x)\,e_x}{m_{xx}} + \frac{(k_y'-k_y)\,e_y}{m_{yy}}
\right] V_{11} \:,
\end{equation}
where $e$ is the electron charge, $A$ is the amplitude of the
electromagnetic wave, $\omega$ is the radiation frequency,
$m_{xx}=m_{\perp}$ and $m_{yy}=(m_{\perp}\cos^2\theta +
m_{\parallel}\sin^2\theta)$ are the effective electron masses in
the channel plane being different from each other due to the
deviation of the channel plane from (001) by the angle $\theta$,
$m_{\parallel}$ and $m_{\perp}$ are the longitudinal and
transverse effective masses in the valley in bulk Si,
$\bm{e}=(e_x,e_y)$ is the unit vector of the light polarization
and $V_{11}$ is the matrix element of electron scattering within
the subband $e1$. While the matrix element\,(\ref{Me1}) is odd in
the wave vector, the absorption probability given by the squared
matrix element is even in $(\bm{k}^\prime-\bm{k})$. Thus, this
type of processes alone does not introduce an asymmetry in the
carrier distribution in $\bm{k}$-space and, consequently, does not
yield an electric current.
%


%
%

Pathways II via states in the  $e2$ subband are sketched in
Figs.\,\ref{figure4_model}d,\,e.  They involve virtual
intersubband optical transitions which in miscut structures are
allowed by selection rules even at normal incidence of
radiation\,\cite{Ando82,Magarill89,Gusev87}.
%
%
%
%
The matrix element of path II indirect optical transitions
has the form
\begin{equation}\label{Me2}
M_{\mathbf{k}'\mathbf{k}}^{(2)} = 2i \frac{eA}{c \hbar}
\frac{m_{zz}}{m_{yz}} \frac{\hbar \omega\,
\varepsilon_{21}z_{21}}{(\hbar\omega)^2 - \varepsilon_{21}^2} e_y
\,  V_{12} \:,
\end{equation}
where
$1/m_{zz}=\cos^2\theta/m_{\parallel}+\sin^2\theta/m_{\perp}$,
$1/m_{yz}=(1/m_{\perp}-1/m_{\parallel})\cos\theta \sin\theta$ is
the off-diagonal component of the reciprocal effective mass
tensor, $\varepsilon_{21}$ is the energy separation between the
subbands, $z_{21}$ is the coordinate matrix element and $V_{12}$
is the matrix element of intersubband scattering. 
Equation\,(\ref{Me2}) shows that this type of indirect transitions
is independent of
$\bm{k}$ and, consequently, also does not lead to an electric
current.

The photocurrent emerges due to the quantum interference of all
virtual transitions considered above. Indeed, the total
probability for the real optical transition $\bm{k} \rightarrow
\bm{k}'$
%
%
is given by the squared modulus of the sum of matrix elements
describing individual pathways
\begin{equation}\label{W_kk}
W_{\bm{k}\prime\bm{k}} \propto
|M^{(1)}_{\bm k^\prime \bm {k}}|^2 + |M_{\bm k^\prime \bm
{k}}^{(2)}|^2 + 2\mathrm{Re}(M^{(1)}_{\bm k^\prime \bm {k}}
M_{\bm k^\prime \bm {k}}^{(2)*}) \,.
\end{equation}
Beside the probabilities of individual processes given by the first and the second term
in the right-hand side of this equation, it contains the interference term.
By using Eqs.\,(\ref{Me1}) and\,(\ref{Me2}) we derive for the latter term
\begin{equation}\label{W_kk}
\mathrm{Re}( M^{(1)}_{\bm k^\prime \bm {k}} M_{\bm k^\prime \bm
{k}}^{(2)*}) \propto (k_x^\prime - k_x) \, i (e_xe_y^* -e_x^*
e_y) \, F(\hbar\omega)\,.
\end{equation}
%
This term is linear in the wave vector and, therefore, it results
in different rates for the transitions to the positive and
negative $k_x^\prime$. This, in turn, leads to an imbalance in the
carrier distribution between $k_x^\prime$  and $-k_x^\prime$,
i.e., to an electric current $j_x$. Such a difference in the real
optical transition rates caused by constructive or destructive
interference of various pathways is illustrated in
Fig.\,\ref{figure4_model}a.
Moreover, the sign of the interference term is determined by the
radiation helicity because
%
%
$i(e_xe_y^* -e_x^* e_y) = \hat{e}_z P_{circ}$, where $\hat{\bm{e}}$ is a unit vector
pointing along the light propagation direction\,\cite{Ivchenko_book,GanichevPrettl_book}.
%
%
Therefore,  the imbalance of the carrier distribution in
$\bm{k}$-space and, consequently, the photocurrent reverse upon
switching  the light
helicity.

Equation\,(\ref{W_kk}) also explains the observed at  low temperature
inversion of the circular photocurrent direction when
the energy separation between the subbands varies from
$\varepsilon_{21}<\hbar\omega$ to $\varepsilon_{21}>\hbar\omega$.
Indeed,
%
the interference term
is proportional to
the function $F(\hbar\omega) \propto 1 / [(\hbar\omega)^2 -
\varepsilon_{12}^2]$, which stems from the matrix element
describing virtual transitions via the $e2$ subband [see
Eq.\,(\ref{Me2})]. In the vicinity of the intersubband absorption
peak, the photocurrent increases drastically and undergoes
spectral inversion.
In real structures
the dependence smooths because of the broadening, but the inversion remains.
At room temperature, when the excited subbands $e2$, $e3$ etc.,
are also occupied in the equilibrium and considerably broaden, the
inversion of the photocurrent does not couple to
$\varepsilon_{21}$ anymore, as observed in experiment.



Assuming the electron  scattering by short-range static defects,
we write for the photocurrent\,\cite{IvchenkoPikus_book,Tarasenko07}
\[
\bm{j} = e \frac{8\pi}{\hbar} \sum_{\bm{k},\bm{k}'}
[\tau_p(\varepsilon_{\bm{k}'}) \bm{v}(\bm{k}') -
\tau_p(\varepsilon_{\bm{k}}) \bm{v}(\bm{k}) ]
[f(\varepsilon_{\bm{k}})-f(\varepsilon_{\bm{k}'})]
\]
\begin{equation}\label{j_gen}
\times 2\mathrm{Re}(M^{(1)}_{\bm k^\prime \bm {k}} M_{\bm
k^\prime \bm {k}}^{(2)*})
\delta(\varepsilon_{\bm{k}'}-\varepsilon_{\bm{k}}-\hbar\omega) \:,
\end{equation}
where $\tau_p(\varepsilon_{\bm{k}})$ is the momentum relaxation
time, $v_x(\bm{k})=\hbar k_x/m_{xx}$ and $v_y(\bm{k})=\hbar
k_y/m_{yy}$ are the velocity components, $\varepsilon_{\bm{k}}$ is
the electron kinetic energy measured from the subband bottom,
$f(\varepsilon_{\bm{k}})$ is the function of equilibrium carrier
distribution in the subband $e1$, and factor 8 in
Eq.\,(\ref{j_gen}) accounts for the spin and valley degeneracy.
Finally, we derive for the circular photocurrent density in
Si-MOSFET structures with a small declination angle $\theta$
\begin{equation}\label{j_final}
j_x = e \tau_p \frac{m_{\parallel}}{m_{yz}} \frac{\langle V_{11}
V_{12} \rangle}{\langle V_{11}^2 \rangle}
\frac{\varepsilon_{21}z_{21} \omega
\eta_x}{\varepsilon_{21}^2-(\hbar\omega)^2} \hat{e}_z I P_{circ}
\:,
\end{equation}
where $\eta_x$ is the channel absorbance for the radiation
polarized along the $x$ axis, $I$ is the radiation intensity, and
the angle brackets stand for averaging over the spacial
distribution of scatterers. Equation\,(\ref{j_final}) describes the
CPGE caused by the free carrier absorption at $\hbar\omega <
\varepsilon_{21}$ when the kinetic energy of photoexcited carriers
is smaller than the intersubband spacing. For electrons with
$\varepsilon_{\bm{k}}
>\varepsilon_{21}$ the momentum relaxation time gets shorter
because of the additional relaxation channel caused by the
intersubband scattering. Consequently, the magnitude of the
current is smaller for $\hbar\omega > \varepsilon_{21}$ (low
$V_g$) than that at $\hbar\omega < \varepsilon_{21}$ (large
$V_g$).  This can be responsible for the observed asymmetry in the
gate voltage dependence of the photocurrent in the intersubband
resonant vicinity, see Fig.\,\ref{figure3_gateLowT}a. At
$\hbar\omega \simeq \varepsilon_{21}$, possible contributions to
the CPGE due to the intersubband optical transitions as well as
the scattering-induced broadening of the absorption peak should
also be taken into account\,\cite{Magarill89}.


%
%

The magnitude of the CPGE current detected in sample 1 for
$\hbar\omega = 8.4$\,meV and $V_{g}$\,=\,3\,V is $J_x/P \sim 1$\,nA/W,
yielding the current density $j_x/I \sim
0.1$\,nA\,cm/W. The same order of magnitude is obtained from
Eq.\,(\ref{j_final}) for the structure with the vicinal angle
$\theta = 9.7^\circ$, the carrier density $N_s= 5 \times
10^{11}$\,cm$^{-2}$ ($V_{g}$= 3 V), the channel width $a = 80$\,\AA
and the structure asymmetry degree $\xi = 10^{-2}$.
%


To summarize, we demonstrate  in experiments on
Si-based structures that the photon helicity-dependent
photocurrents can be generated in low-dimensional 
semiconductors even with vanishingly small spin-orbit
interaction. The mechanism of the photocurrent formation is based
on the quantum interference of different pathways contributing to
the radiation absorption.

\acknowledgments We thank E.L.\,Ivchenko, V.V. Bel'kov,
L.E. Golub and S.N. Danilov.
The financial support from the DFG and the RFBF is gratefully acknowledged.

\newpage

\end{document}